# Comment on "Large energy gaps in $CaC_6$ from tunneling spectroscopy: Possible evidence of strong-coupling superconductivity"

N. Emery[1], C. Hérold[1], S. Cahen[1], H. Rida[1], J. F. Marêché[1], P. Lagrange[1], G. Lamura[1,2]

[1]*Laboratoire de Chimie du Solide Minéral-UMR 7555, Université Henri Poincaré Nancy I, B.P. 239, 54506 Vandœuvre-lès-Nancy Cedex, France.*

[2]*CNR-INFM Coherentia and Department of Physics, University of Naples "Federico II", 80125 Naples, Italy.*

*PACS*: 74.62.Bf; 74.70.Ad.

## Abstract

Recently, C. Kurter *et al.*, [Phys. Rev. B **76**, 220502(R) (2008)] report on tunnelling measurement on bulk $CaC_6$. Their samples have been obtained by using the molten alloy method described by Emery *et al.* [Phys. Rev. Lett. **95**, 087003 (2005)] with natural, single-crystal, graphite flakes. They claim that the final product of the intercalation process consist of $CaC_6$ crystals. In this comment we show why this claim is ambiguous or even wrong when any kind of graphite sample is intercalated by using the molten alloy method. At the same time, because of the lamellar structure of $CaC_6$, the inappropriate terminology used to describe the sample *does not* question the interesting scientific results and the conclusions of the paper for any reason.

In a recent Rapid Communications Letter, C. Kurter *et al*., [1] report on tunneling measurements in both superconductor–insulator–normal metal (SIN) and (SIS) junctions in $CaC_6$, a newly discovered superconductor that belongs to the family of graphite intercalation compounds (GICs). They claim that the samples used in their measurements were *crystals* synthesized with the "alloy method", originally devised by the Nancy group some years ago [2-4]. However, this assumption is incorrect, as we explain in detail below. Graphite can be available as Highly Oriented PyroGraphite (HOPG) which is polycrystalline c-axis oriented or small single crystals like natural Madagascar graphite. A closer inspection of the wide literature on GICs reveals that the intercalation process even starting with ***graphite single crystals always gives rise to c-axis oriented polycrystals***.

Upon GICs formation, intercalate atoms penetrate the single crystal graphite layers without preferential directions and they are ordered following one of the six graphite senary axis. Since these directions are equivalent, the intercalate will form single-crystal, multiple-twinned domains, rotated by 60° with respect to each other, and peculiar to GICs. At the end of the intercalation process, the crystal structure consists therefore of two overlapping sublattices, that of the graphene layers and the one of the intercalated atoms. Even though the former remains a single crystal, this is not the case for the latter. As a result, the final products of the intercalation process on pristine graphite single crystals are unavoidably c-axis oriented polycrystals [5].

The study of the intercalate sublattice has been a hot topic for several decades in the GICs community, with great attention being reserved to first stage compounds as $KC_8$, $RbC_8$ and $CsC_8$ [6-8]. Among these, the well known case of $KC_8$ obtained from a graphite single crystal clearly illustrates the relation between the graphite and the intercalate atoms sublattices [6]. The unit cell of this GIC is orthorhombic (Figure 1), so that the c-axis diffraction pattern of a true single crystal should never show a hexagonal symmetry. Nevertheless, all X-ray experiments lead always to the presence of a hexagonal symmetry. To understand this, one has to consider the resolved intercalate sublattice, which consists of coexisting orthorhombic cells derived from each others by a rotation of 60° around the graphite c-axis. From the width of the *hk0* reflections the domain dimensions have been estimated in a few tens of nanometres, at least in the intercalated layers. In the c-axis direction the existence of several stacking faults is highly probable [5,6].

As $KC_8$, $RbC_8$ also possesses an orthorhombic unit cell [7]. As in the previous case, the c-axis diffraction pattern exhibits a hexagonal symmetry, due, for the same reasons, to six coexisting orthorhombic cells. On the contrary, in the case of $CsC_8$, the unit cell is hexagonal [8], because the c-axis intercalate atoms stacking is different from the previous ones. Since the hexagonal symmetry already exists for the $CsC_8$ unit cell, it is of course impossible to highlight, as previously, the presence of six twinned domains, rotated by 60° with respect to each other. Nevertheless, it is absolutely logical to consider that the intercalation process remains the same wathever the metal to intercalate. In conclusion, because of this lack of preferential ordering directions during the synthesis, the presence of multiple-twinned domains inevitably appears. Thus, in the case of a rhombohedral GIC compound as $CaC_6$, it is clear that the same phenomenon turns up.

The polycrystalline nature of the intercalate sublattice is at the origin of the largely studied case of liquid-solid like phase transitions in some second stage GICs [9-12]. In heavy alkali metals GICs, at room temperature the intercalate atoms are in a 2D liquid-like phase (absence of interlayer correlations) which is incommensurate with the graphite sublattice. As the temperature decreases and the interlayer correlations become important (2D-3D crossover), the intercalate atoms order through a liquid-solid like transition between 100 and 170 K in islands of tens of nanometres in size, as in case of K, Rb and Cs second stage GICs [9-12].

Finally, we recall that several groups describe the $CaC_6$ bulk samples, synthesized through the method described in ref. [2-4], as golden in colour [13,14]. If extremely pure calcium is used (dendritic calcium 99.99% from Sigma-Aldrich), the colour of the samples obtained by the molten Li-Ca alloy method [2-5] is a light shining silver. The golden reflex reported for the $CaC_6$ samples could be due to several reasons: a slight degradation of the sample surface, the presence of other phases such as $LiC_6$ [15-18], depending on thermodynamic conditions during the synthesis process of the Li-Ca alloy method [2-4]. Careful attention of the X-ray diffraction datas should always be used in order to exclude the presence of small amounts of other GICs. X-ray diffraction (XRD) is not sufficient to state on the presence of Li-based inclusions because of its negligible scattering amplitude but they can be revealed by nuclear microprobe analysis (NMA) [19]. The combined use of different techniques like XRD and NMA helps of course to characterize the purity of GIC samples.

**Figure captions**

**Fig. 1**. $KC_8$. This stoichiometry of $KC_8$ corresponds to an AA stacking of the successive graphene planes; i.e. all the carbon atoms in two successive layers are superimposed. Intercalate metal atoms in each layer are located in one out of four prismatic hexagonal sites denoted by (– 0, 8 ¼, , ½, 7 ¾) that gives rise to three different orthorhombic unit cells, each rotated by 60° with respect to the others.

**Fig. 2**. (Color online) Bulk $CaC_6$ and $LiC_6$ samples.

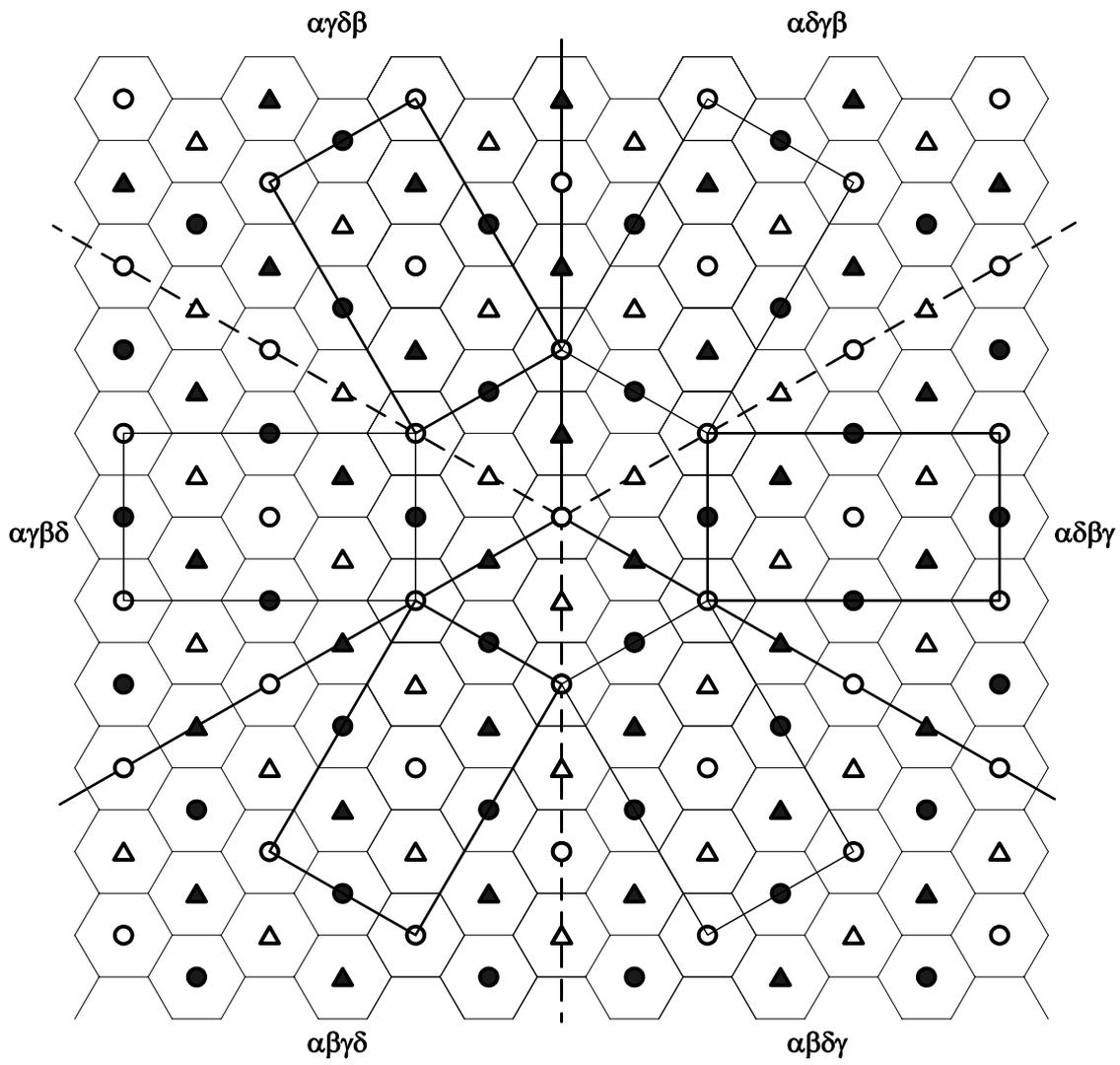

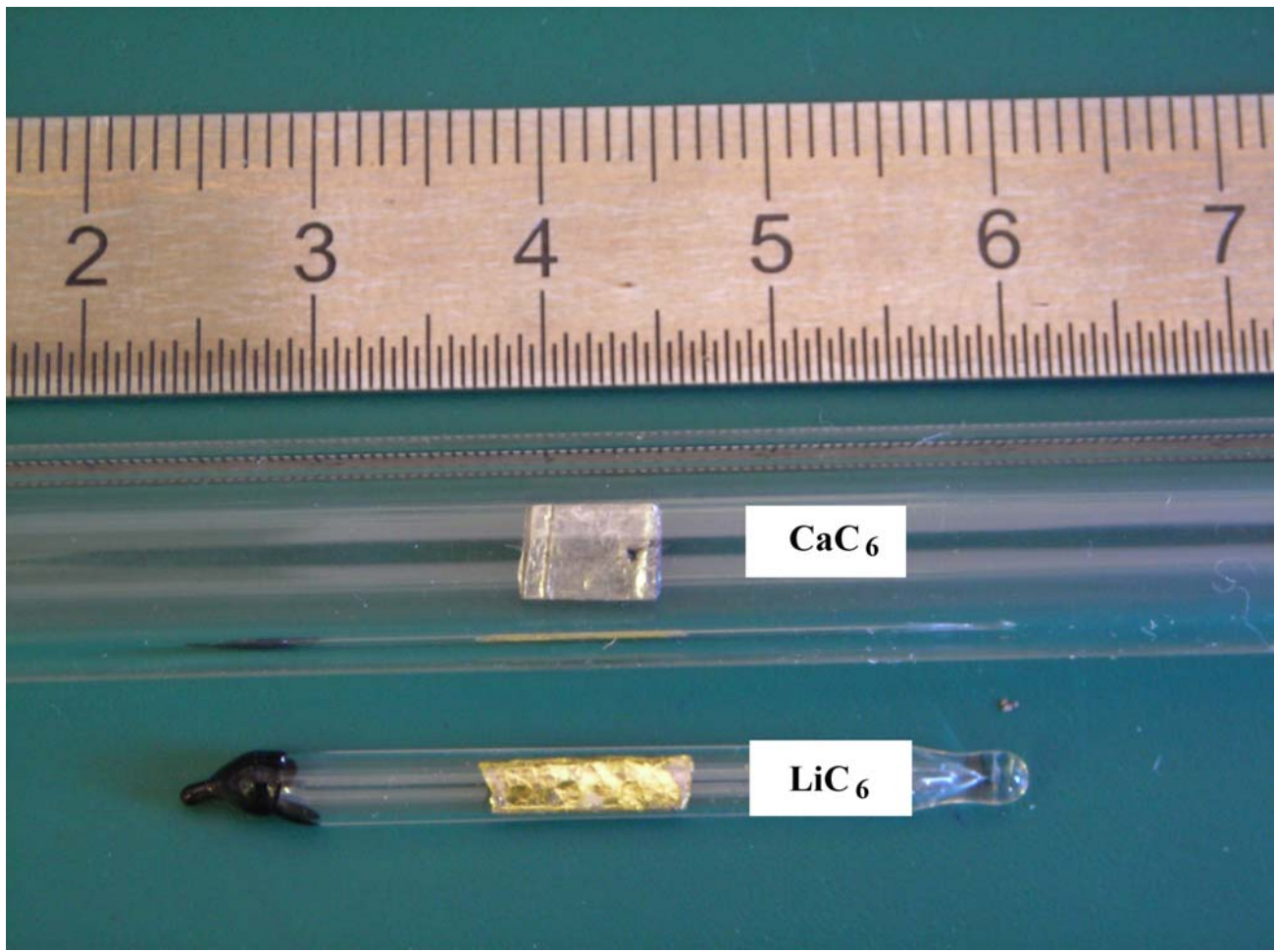